\documentclass[a4paper,12pt]{article}
\usepackage[utf8]{inputenc}

\usepackage[T1]{fontenc}
\usepackage{pifont}
\usepackage{amsmath}
\usepackage{amssymb}
\usepackage{amsthm}
\usepackage{amscd}

\usepackage{caption}
\usepackage{subcaption}
\usepackage{xcolor}
\definecolor{light-gray}{gray}{0.80}
\definecolor{dark-gray}{gray}{0.25}

\usepackage{tikz}
\usetikzlibrary{decorations.markings,positioning,fit,arrows,decorations.pathreplacing}
\usetikzlibrary{calc, patterns}

\usepackage{pgfplots}
\pgfplotsset{compat=1.5}

\newtheorem{theorem}			     {Theorem}

\newcommand{\Z}{\mathbb{Z}}

\newcommand{\minsite}{\tikz[baseline=-0.5ex]{\node[circle,draw]{}; \node[]{$-$};}}
\newcommand{\plussite}{\tikz[baseline=-0.5ex]{\node[circle,draw]{}; \node[]{$+$};}}

\title{Intermediate model between Majority voter PCA and its Mean Field model}

\author{Jean Bricmont\footnote{Partially supported by the Belgian IAP program P7/18.}
\\{\normalsize Universit\'{e} catholique de Louvain, IRMP},\\{\normalsize chemin du Cyclotron 2, L7.01.03}\\ {\normalsize B-1348 Louvain-la-Neuve (Belgium), \texttt{Jean.Bricmont@uclouvain.be}}
\vspace{8mm} \\ Hanne Van Den Bosch
	\\{\normalsize Universit\'{e} catholique de Louvain, IRMP},\\{\normalsize chemin du Cyclotron 2, L7.01.03}\\ {\normalsize B-1348 Louvain-la-Neuve (Belgium)}
\\  {\normalsize and Pontificia Universidad Cat\'{o}lica de Chile} \\
{\normalsize Av. Vicu\~{n}a Mackenna 4860} \\
{\normalsize Santiago (Chile), \texttt{hannevdbosch@fis.puc.cl}} 
}
\date{October 2, 2013}

\begin{document}

\maketitle

\begin{abstract}
 Probabilistic Cellular Automata (PCA) are simple models used to study dynamical phase transitions. 
There  exist mean field approximations to PCA that can be shown to exhibit a phase transition.
We introduce a model interpolating between a class of PCA, called majority voters, and their corresponding mean field models. 
Using graphical methods, we prove that this model undergoes a phase transition.
\end{abstract}

\section{Introduction}
We consider probabilistic cellular automata defined on $\Z^2$ or on a finite box in $\Z^2$. 
Each cell in this space can be in two possible states, \plussite and \minsite.
Each cell interacts with a finite and odd number of neighbors (possibly including itself).
Two examples of neighborhoods are shown in figure \ref{fig : neighbors}.
All sites update simultaneously according to some transition probabilities.
For the majority voter PCA, each site becomes the state of the majority of its neighbors 
with probability $1-\epsilon$ and the opposite with probability $\epsilon$, for some error rate $\epsilon \in [0,0.5]$. 
One naturally associates an operator $P$ on measures on configurations to these transition probabilities.
When iterated starting from some initial measure $\mu$, we obtain a sequence of measures $\mu, P \mu, P^2 \mu, \hdots$.

A measure $\mu$ is  \emph{invariant} for a PCA $P$ if $\mu =  P \mu $.
If there is an invariant measure $\tilde \mu$ such that $\lim_{n \to \infty} P^n \mu = \tilde \mu$ (in the sense of weak convergence) for any initial measure $\mu$, 
the PCA is  \emph{ergodic}.
Dobrushin's criterion \cite{Maes} shows that this is the case 
for any majority voter PCA when the error rate is sufficiently close to $0.5$. 
Toom's theorem \cite{Toom_proof}
shows that some majority voters with an asymmetric neighborhood are non-ergodic for sufficiently small error rates.
The transition between an ergodic and a non-ergodic regime that these models undergo as $\epsilon$ varies is called a \emph{phase transition}.
For the symmetric majority voter with five neighbors,
 numerical simulations \cite{Kozma} suggest that a similar phase transition occurs, but there is no proof of it.

\begin{figure}
\caption{}
\begin{subfigure}[b] {0.4 \textwidth}
\begin{center}

  \begin{tikzpicture}[scale = 0.6,
	    plus/.style = {draw, shape = circle, color = light-gray, fill = white, scale = 1},
	     neighbor/.style = {ultra thick, gray, fill = light-gray }
]		      
\draw [neighbor] (-0.5, -0.5) -- ++ (2,0)-- ++ (0,1) -- ++ (-1,0) -- ++ (0,1)-- ++ (-1,0) -- ++ (0,-2) ;
\draw [neighbor] (3.5, 0.5)-- ++ (0,-1)--++ (1,0) --++ (0,1)-- ++ (1,0)-- ++ (0,1) 
		    -- ++ (-1,0) -- ++ (0,1)-- ++ (-1,0) -- ++ (0,-1) 
		    -- ++ (-1,0) --++ (0,-1) -- ++ ( 1,0);
\foreach \x in {-1,...,6}
 \foreach \y in {-1,...,3}
 \node [plus] at (\x,\y){};
\node  [plus, thick, color = black, fill = white] at (0,0){};
\node  [plus, thick, color = black, fill = white] at (4,1){};
    \end{tikzpicture}
\end{center}
\caption{Neighborhood for Toom's majority voter (North-East-Center) 
	      and for the symmetric voter with five neighbors.
	      For the first one, a phase transition has been proven, but not for the second one.} \label{fig : neighbors}
\end{subfigure}
\begin{subfigure}[b] {0.5 \textwidth}

\begin{center}

  \begin{tikzpicture}[scale = 0.6,
	plus/.style = {draw, shape = circle, color = light-gray, fill = white, scale = 1},
        neighbor/.style = {ultra thick, gray, fill = light-gray }
]
\draw [neighbor](-3.5,3.5)--(3.5,3.5)--(3.5,-3.5)--(-3.5,-3.5)--cycle;
 \foreach \x in {-5,...,5}
  \foreach \y in {-4,...,5}
{\node (a\x\y) [plus ] at (\x,\y) {};}

\node (O) [plus, thick, color = black, fill = white] at (0,0){};
\draw[->][ thick] (O) to[bend right] (a-3-2);
\draw[->][ thick] (O) to[bend right] (a12);
\draw[->][ thick] (O) to[bend right] (a2-1);
\draw[->][ thick] (O) to[bend left] (a2-1);
\draw [->][ thick] (O) to[bend right] (a-33);
\node (l)[shape = rectangle] at (0,-2.5) {$l$};
\draw[->][thick] (l) to (-3.5,-2.5); 
\draw[->][thick] (l) to (3.5,-2.5); 
\node (g) at (1.8,3)[ align = left] {$l^2 = \gamma$ sites};
\node (R)[shape = rectangle] at (0,4.5) { $R$};
\node[above = 0cm of R] {lattice size};
\node[below = -0.2 cm of l]{interaction range};
\draw[->][thick] (R) to (-5.5,4.5); 
\draw[->][thick] (R) to (5.5,4.5); 

\end{tikzpicture}
\end{center}

\caption{Notations for the intermediate model: the central site picks $b$ neighbors randomly within a square of side $l$.} 
\label{fig : intermediate}
\end{subfigure}
\end{figure}

\section{Mean Field model}
Since it appears hard to prove anything for PCA not covered by Toom's theorem,
a mean field approximation has been considered.
In this approximation, defined in \cite{Balister}, 
the interaction range becomes equal to the size of the system.  Consider a box of side $R$ in $\Z^2$.
Instead of interacting with $b$ fixed neighbors, each site picks at each time step $b$ random sites, uniformly in the whole box.
It then updates to the majority of these sites with probability $1-\epsilon$ and the opposite with probability $\epsilon$.
In this case, only one variable, e.g. the density of \plussite, is sufficient to describe the state of the model.
Moreover, the transition probabilities for the density can be easily computed and the evolution tends to a deterministic one 
when $R$ tends to infinity.
The mean field model undergoes a phase transition for any majority voter model, see  \cite{Balister}.
However, we don't know, in general, 
when mean field models are good approximations to models with finite range interactions. 
This motivates the construction of models interpolating between the majority voter PCA and its corresponding mean field model.

We are looking for some analogue of Kac potentials in statistical mechanics \cite{Thompson}.
In these models, the interaction range is a parameter of the potential. 
It is known that the free energy of these models (in the thermodynamic limit) tends to the mean field free energy, when the
range of the interaction tends to infinity.

\section{Intermediate model}
We now consider a PCA where the interaction range $l$ is a parameter.
As shown in figure \ref{fig : intermediate},
at each time step, each site chooses an odd number $b$ of neighbors in a square of side $l$ centered on this site.
It then takes the state of the majority of the chosen states with probability $1 - \epsilon$ and the opposite with probability $\epsilon$.
The points in space-time that do not follow the majority rule will be called error sites.
Note that, when $l$ is equal to $R$, the side of the box, 
this model is the mean field model (assuming periodic boundary conditions).

For any fixed value of $l$, this is a well-defined probabilistic cellular automaton, even when $R$ becomes infinite.
One may want to study this model for large, but fixed, interaction range $l$.
However, this appears to be  difficult.
As a further simplification, we consider an intermediate model, where the interaction range $l$ changes at each time-step.
More precisely, 
we fix some time $T>0$ and let $\gamma_t = l_t^2$ 
be the number of sites at time $T-t-1$ among which a site at time $T-t$ chooses its neighbors.

\begin{theorem}
 The intermediate model defined above, with $\gamma_t= g p ^t$,
is non-ergodic for sufficiently small values of $\epsilon$ and sufficiently large values of $g$ and $p$.
\end{theorem}

\noindent {\bf Remarks} 

\noindent 1. The model is artificial, but the proof of the Theorem is instructive. It starts with a rather standard construction (given in the next section) 
of graphs that are ``responsible" for a site being  in the state \minsite 
at some time $T$ when the initial configuration is \plussite. The probability of these graphs is small because of two different effects: either there are lots of error sites, each of which has probability $\epsilon$,
or, at some moments of time, the same site is chosen several times, which reduces the ``entropy" of the graph: since a site is chosen with a probability $\gamma_t ^{-1}$, among $\gamma_t $ sites, if the same site is chosen twice,
we get an extra factor $\gamma_t ^{-1}$. This is expressed in more detail in equation
 (\ref{eq : Prob}) below. However, because of the combinatorics of the graphs, we cannot complete the proof without letting $\gamma_t $ grow exponentially with $t$.
 
\noindent 2. The model is trivially ergodic for $\epsilon= \frac{1}{2}$ (since, then, each site becomes \plussite or   \minsite  independently of the past); for  $\epsilon$ close to $\frac{1}{2}$, one may use Dobrushin's criterion
adapted to PCA (see \cite{Maes}) to show that the model is still ergodic, which proves 
 that the intermediate model undergoes a phase transition.

\noindent 3. We did some numerical simulations, see table \ref{tab : simulations}, indicating that our model for fixed $\gamma = l^2$, with $l=5$, undergoes a phase transition when $\epsilon$ varies, but we are unable to prove it.

\begin{table}
\caption{Mean density of plus-sites in simulations after 1000 time steps started from density $1$, 
with $b=5$, $l=5$, on a grid of size $R=120$, for different values of $\epsilon$. 
In all cases, the density was observed to reach its equilibrium value in less than 100 time steps.}
\label{tab : simulations}

 \begin{center}
\begin{tabular}{|r |llll|}
\hline
$\epsilon$ & 0.15 & 0.175 & 0.2 & 0.225\\ \hline
density & 0.782 & 0.582 & 0.500 & 0.500 \\
\hline
 \end{tabular}
 \end{center}

\end{table}

\section{Proof}

The idea of the proof  is to use graphical methods to bound the probability that a given site, say the origin, is in state \minsite 
at some time $T$ when the initial configuration is \plussite, which will be noted $P(\omega_T^0 = -)$.
If we obtain a bound on this quantity which is independent of $T$ and smaller than $\frac{1}{2}$ for small values of $\epsilon$, 
this proves the theorem.

\subsection{Constructing graphs}
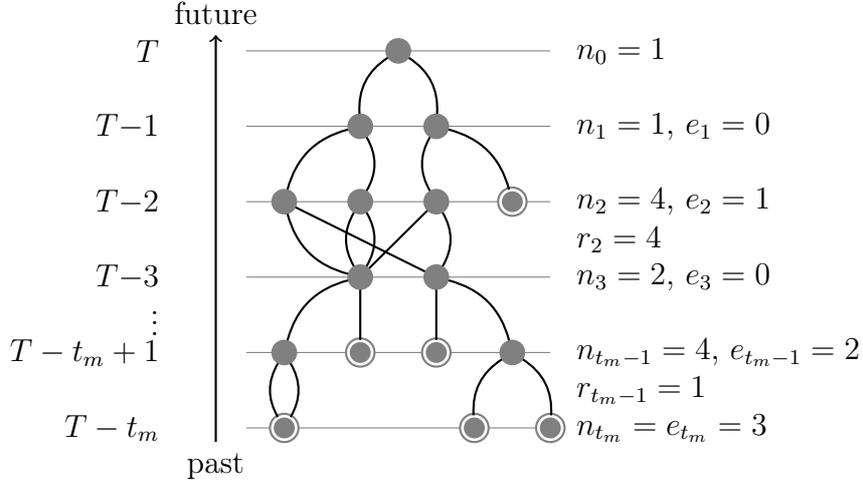
\begin{figure}

\begin{center}
 
\begin{tikzpicture}[scale = 1,site/.style = {draw, thick, shape = circle, color = gray, fill = gray, scale = 0.8},
	     error site/.style = {draw, thick, double, double distance = 1pt, shape = circle, color = gray, fill = gray, scale = 0.8}
	     ]
\foreach \t in {0,...,5}
  \draw[thin, color = gray] (-2,-\t)--(2,-\t);
\node (futur) [] at (-2.4,0.5){future};
\node (passé) [] at (-2.4,-5.5){past};
\draw [->][thick] (passé)--(futur);
\foreach \t in {1,...,3}
\node () [anchor = east] at (-3,-\t){$T-$\t} ;
\node () [anchor = east] at (-3,0) {$T$} ;
\node () [anchor = east] at (-3,-3.5) {$\vdots$};
\node () [anchor = east] at (-3,-4) {$T-t_m + 1$};
\node () [anchor = east] at (-3,-5) {$T-t_m$};
\node () [anchor = west] at (2.2,0) {$n_0 = 1$} ;
\node () [anchor = west] at (2.2,-1){$n_{1} = 1$, $e_{1}= 0$} ;
\node () [anchor = west] at (2.2,-2){$n_{2}=4$, $e_{2}=1$} ;
\node () [anchor = west] at (2.2,-3){$n_{3}=2$, $e_{3}= 0$} ;
\node () [anchor = west] at (2.2,-4) {$n_{t_m-1} = 4$, $e_{t_m-1}= 2$};

\node () [anchor = west] at (2.2,-5) {$n_{t_m} = e_{t_m} = 3$};
\node (A) [site ] at (0,0) {};
\node (B) [site ] at (-0.5,-1){};
\node (C) [site ] at (0.5,-1){};
\foreach \x in {-1,...,1}
\node (D\x) [site] at ({\x-0.5},-2){};
\node (De) [error site] at (1.5,-2){};
\node (E1) [site ] at (-0.5,-3){};
\node (E2) [site ] at (0.5,-3){};
\node (F1) [site ] at (-1.5,-4){};
\node (F2) [error site ] at (-0.5,-4){};
\node (F3) [error site ] at (0.5,-4){};
\node (F4) [site ] at (1.5,-4){};
\node (G1) [error site ] at (-1.5,-5){};
\node (G2) [error site ] at (1,-5){};
\node (G3) [error site ] at (2,-5){};
\foreach \from / \to in {A/B, B/D-1, C/D1, D-1/E1, D0/E1, E1/F1, F1/G1, F4/G2} \draw[thick] (\from) to[bend right] (\to);
\foreach \from / \to in {A/C, B/D0, C/De, D0/E1, D1/E2, E2/F4, F1/G1, F4/G3} \draw[thick] (\from) to[bend left] (\to);
\foreach \from / \to in { D-1/E2, D1/E1, E2/F3, E1/F2} \draw[thick] (\from) to (\to);
\node () [anchor = west] at (2.2,-2.5) {$r_2 = 4$};
\node () [anchor = west] at (2.2,-4.5) {$r_{t_m-1} = 1$};
\end{tikzpicture}

\end{center}

\caption{Example of a graph $G$ for the case with three neighbors : 
$n_t$ counts the number of sites, $e_t$ error sites, and $r_t$ the number of extra lines.}
\label{fig : graph}
\end{figure}

The history of site $0$ at time $T$ can be represented by a graph $H$ in space-time.
From the point $(0,T)$, $b$ lines leave towards the sites chosen at time $T-1$ and so on  (these sites will be the vertices of the graph, and the lines its edges).
It is important to notice that the $b$ sites are chosen independently of each other, 
so it is possible that the same site is chosen twice or more.
If the origin is in state \minsite at time $T$, 
it is either an error site, or the majority of the sites it chose are in state \minsite too. 
In this case we can choose $k = \frac{b+1}{2}$ ($b$ is odd)  lines of $H$ leaving from $(0,T)$ that are incident on sites in state \minsite.
Iterating this procedure we can extract a subgraph $G$ of $H$ such that all sites of $G$ are \minsite, 
exactly $k$ lines leave from each site which is not an error site and no lines leave from error sites.
Figure \ref{fig : graph} shows an example of such a graph for the case of $b= 3$ neighbors.
Since all sites at time $0$ are in state \plussite, such a graph always exists
(it is not unique in general).
We thus have 
\[
 P(\omega_T^{0}= - ) \leq \sum \limits_{\substack{\\ \text{graphs G}}} \text{probability of } G.
\]
The sum runs over all graphs that can be associated to a space-time history of the model in the above described manner.

\subsection{Probability of graphs}

Each error site in a graph adds a factor $\epsilon$ to the probability of a graph.
Another contribution to the probability of graphs comes from the lines.
For fixed $T$,
$\gamma_t$ is the number of sites at time $T-t-1$ on which  the state of a site at time $T-t$ may depend.
In the graph $H$, the probability that the lines leaving from a given point $(x,T-t)$ are entering  sites $s_1, s_2, \hdots, s_b$ 
(in that order)
equals $\frac{1}{\gamma_t^b}$.
We would like to compute the probability that, if the graph $G$ contains $(x,T-t)$, 
the lines leaving from it enter the sites $p_1, p_2, \hdots, p_k$.
Therefore we multiply the former probability by $\binom{b}{k} \gamma_t^{b-k}$,
 which counts the number of sequences $s_1, s_2, \hdots, s_b$ containing $p_1, p_2, \hdots, p_k$, the factor $\gamma_t^{b-k}$ coming from the sum over the $b-k$ sites not
 in $\{p_1, p_2, \hdots, p_k\}$.
For simplicity, we will use the notation $C\equiv \binom{b}{k}^\frac{1}{k} $. So, we have a factor $\epsilon$ for each error site and a factor $(\frac{C}{\gamma_t})^k $
for each site that is not an error site.

For a fixed graph $G$, let $n_t$ be the number of sites at time $T-t$, and $e_t$ the number of error sites. Thus, the number of sites
at time $T-t$ that are linked by  lines  to  sites at time $T-t-1$ equals $n_t - e_t$.
Denote by $t_m$  the height of the graph, so that $n_{t_m} = e_{t_m}$ (see also figure \ref{fig : graph}).
Combining the contribution from error sites and edges, we have:
\begin{equation} \label{eq : Prob}
P(\omega_T^{0}= - ) \leq \sum \limits_{\text{graphs G}} 
	  \prod_{t = 0}^{t_m (G)} \epsilon^{e_t} \left( \frac{C}{\gamma_t} \right)^{k (n_t - e_t)}.
\end{equation} \label{eq : kappa}

\subsection{Graph Combinatorics}

First, we define two graphs to be \emph{equivalent } if they differ only by the  spatial positions of their vertices. 
Independently of the shape of the graph, a site at time $T-t-1$ has at most $\gamma_t$ spatial positions 
once we fix the positions of the sites at time $T-t$  with which   it is connected. Thus,

\[
P(\omega_T^{0}= - ) \leq \sum \limits_{\substack{\text{non-equivalent }  \\ \text{graphs G}}} 
	  \prod_{t = 0}^{t_m (G)} \epsilon^{e_t}C^{k (n_t - e_t)} \left( \frac{1}{\gamma_t} \right)^{k (n_t - e_t)-n_{t+1}} .
\]

We would now like to obtain an upper bound on the number of different graphs with a fixed number of sites and error sites at each time.
Starting from the top site, we have to assign an ``arrival site'' to each of the $k$ lines leaving from it.
Since we already summed over different spatial positions, the sites at time $T-1$ are all equivalent.
We can inductively  assign arrival sites at each time step. 
We will denote by $\kappa^n_m$ the number of ways to arrange lines between $n$ sites at time $T-t$ and $m$ sites at time $T-t-1$.
In what follows, we will use the bound:

\begin{equation} \label{eq : kappa}
 \kappa^n_m \leq \frac{(kn)! \binom{kn-1}{m-1}} {m !}.
\end{equation}

\begin{figure}

 \begin{tikzpicture}[scale = 1, site/.style = {draw, thick, shape = circle, color = gray, fill = gray, scale = 1},]
  \node (A) [site ] at (0,0) {};
   \node (B) [site ] at (2,0) {};
   \node () [ ] at (3.5,0) {$\hdots$};
   \node (C) [site ] at (5.5,0) {};
\foreach \x in {1,..., 6}
 \node (a\x)[] at ({\x*0.75-1.5},-1) {$l_\x$};
 \node () [ ] at (3.6,-1) {$\hdots$};
 \node (a7)[] at (4.75,-1) {};
\node (a8)[] at (5.5,-1) {$l_{kn-1}$};
\node (a9)[] at (6.25,-1) {$l_{kn}$};
\foreach \x in {1,..., 6}
 \node (b\x)[] at ({\x*0.78-1.5},-3.5) {$l_{\sigma(\x)}$};
 \node (b7)[] at (4.5,-3.5) {};
\node (b8)[] at (5.3,-3.5) {$l_{\sigma(kn-1)}$};
\node (b9)[] at (6.4,-3.5) {$l_{\sigma(kn)}$};
\coordinate(perm) at (2.7,-2);
\foreach \x in {1,2,3} \draw[gray] plot[smooth] coordinates {(A)(a\x)(perm)};
\foreach \x in {4,5,6} \draw[gray] plot[smooth] coordinates {(B)(a\x)(perm)};
\foreach \x in {8,9} \draw[gray] plot[smooth] coordinates {(C)(a\x)(perm)};

\foreach \x in {1,...,6,8,9} \draw[gray] plot[smooth] coordinates {(perm)(b\x.north)(b\x.south)};

\foreach \x in {1, 4, 5, 8} \draw[->, thick] ([yshift = -15pt]b\x.center) to (b\x);
\node[fit= (b1)(b2)(b3)](group1){};
  \draw[line width=1pt,decorate,decoration={amplitude=7pt,brace,mirror}]
 ([yshift = -5pt]b1.south) -- ([yshift = -5pt]b3.south);
  \node[below= 0.5cm of group1,anchor=center,site,label = below:$p_1$]{};
\node[fit= (b4)](group2){};
\node[below=0.5 cm of group2,anchor=center,site,label = below:$p_2$]{};
\node[fit= (b5)(b6)(b7)](group3){};
  \draw[line width=1pt,decorate,decoration={amplitude=7pt,brace,mirror}]
 ([yshift = -5pt]b5.south) -- ([yshift = -10pt]b7.south);
  \node[below=0.5cm of group3,anchor=center, site,label = below:$p_3$]{};
\node[fit= (b8)(b9)](group4){};
  \draw[line width=1pt,decorate,decoration={amplitude=7pt,brace,mirror}]
 ([yshift = -5pt]b8.south) -- ([yshift = -5pt]b9.south);
  \node(X)[below=0.5 cm of group4,anchor=center,site, label = below:$p_m$]{};
\node[draw, fill = white, minimum height = 1.2cm, minimum width = 5 cm]() at (2.7,-2){    $(k n)!$ permutations  $\sigma$  };
\node[anchor = west](x) at (7.5,-4){$\binom{kn-1}{m-1}$ positions for arrows};
\node[anchor = west, text width = 5cm](y) at (7.5,-4.9){$m$ indiscernible sites : $m !$ permutations};
\node[anchor = west] (z) at (7.5, 0){$n$ sites};
 \node (A) [site ] at (0,0) {};
   \node (B) [site ] at (2,0) {};
   \node (C) [site ] at (5.5,0) {};
\draw[->](6,0)--(z);
\draw[->](6,-4.9)--(y);
\draw[->](6.8,-4)--(x);
 \end{tikzpicture}
\caption{Majoration for $\kappa_m^n$ : the lines are classified in the order of the sites they enter, 
an arrow points to the first line of each arrival site. } 
 \label{fig : combi}
\end{figure}
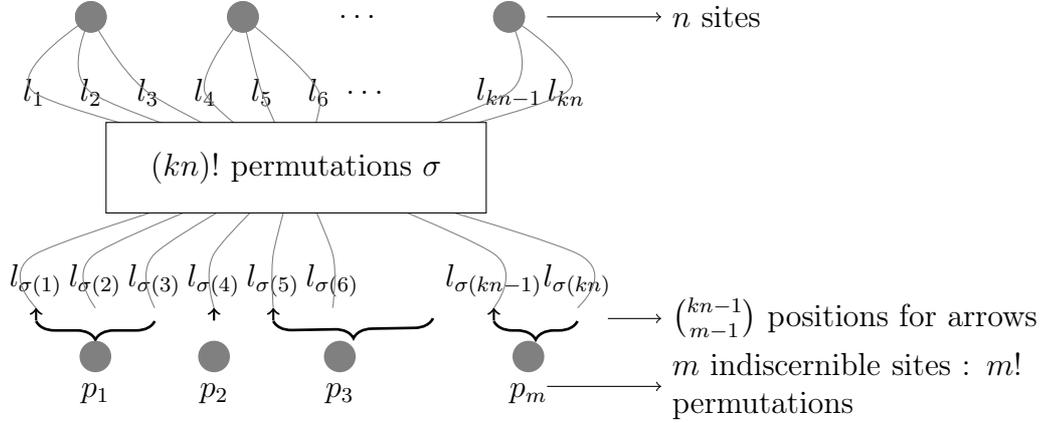

This can be seen, see figure \ref{fig : combi}, by giving arbitrary names $s_1, \hdots, s_m$ to the arrival sites, 
permuting the lines so that the ones assigned to $s_1$ come first, then the ones assigned to $s_2$ and so on.
There are $(kn) !$ of these permutations.
Afterwards, we have to pick the first line assigned to each of the $m-1$ sites $s_2, s_3, \hdots, s_m$ (indicated by arrows in figure \ref{fig : combi}), which gives the binomial factor.
Finally, the names given to the arrival sites were arbitrary, so we can divide by $m !$.
There is still  a combinatorial factor counting the number of ways to 
pick $e_t$ error sites between  $n_t$ sites.
This gives an additional factor $\binom{n_t}{e_t}$. So,  we get:

\begin{equation} \label{eq : sum_tm}
P(\omega_T^{0}= + ) \leq \sum \limits_{t_m = 0}^T 
 \sum\limits_{ (n_t,e_t)_{t \leq t_m}  } 
	\prod \limits_{t = 0}^{t_m - 1} 
		  \binom{n_t}{e_t}
		\kappa_{n_{t+1}}^{n_t-e_t} 
		\epsilon^{e_t} C^{k(n_t -e_t)} 
		\left( \frac{1}{\gamma_t} \right)^{k (n_t - e_t)-n_{t+1}}
		  \epsilon^{n_{t_m}},
\end{equation}
where the sum is over sequences $(n_t, e_t)_{t \leq t_m}$ such that $0< n_{t+1} \leq k (n_t - e_t)$, $e_{t_m}= n_{t_m}$.
The constraint $0< n_{t+1} \leq k (n_t - e_t)$ comes from the fact that there are $k$ lines coming out of  $  (n_t - e_t)$ sites at time $T-t$, so
they can reach at most $k (n_t - e_t)$ sites at time $T-t-1$.

\subsection{Summing over sequences with fixed $t_m$}
The idea is to sum first over the different possibilities for the lowest points of the graph 
and to write the result as a sum over shorter graphs with a new value of $\epsilon$.
Define $r_t= k (n_{t}- e_{t}) - n_{t+1}$ and let
\begin{align}  
S_{\gamma_t} (t_m, \epsilon, \epsilon_0) \equiv & \sum \limits_{ (n_t,e_t)_{t \leq t_m}  } 
	      \prod \limits_{t = 0}^{t_m - 1} 
		\frac{
		\binom{n_t}{e_t}
		\kappa_{n_{t+1}}^{n_t-e_t} 
		\epsilon^{e_t} C^{k(n_t -e_t)}}
		 {\gamma_t^{r_t}}
		  \epsilon_0^{n_{t_m}} 			
  \nonumber    \\
		= & \sum \limits_{ (n_t,e_t)_{t \leq t_{m-1}}  } 
		\prod \limits_{t = 0}^{t_m - 2} 
		\frac{
		\binom{n_t}{e_t}
		\kappa_{n_{t+1}}^{n_t - e_t} 
		\epsilon^{e_t} C^{k(n_t -e_t)}}
		 {\gamma_t^{r_t}} 
 \nonumber  \\
 & \quad \quad \cdot
		  \epsilon^{e_{t_m-1}}\binom{n_{t_m-1}}{e_{t_m-1}}  
\nonumber   \\
 & \quad \quad \cdot
		  C^{k(n_{t_m-1} -e_{t_m-1})}
		\sum \limits_{n_{t_m} = 1}^{k( n_{t_m-1}-e_{t_m-1})}
		 \frac{
		      \kappa_{n_{t_m}}^{n_{t_m-1}-e_{t_m-1}}
		    \epsilon_0^{n_{t_m}}}
		    {\gamma_{t_{m-1}}^{r_{t_m-1}}}. 
\label{eq : def_S}
\end{align}
If we set $\epsilon_0 = \epsilon$, this is exactly the inner sum in equation (\ref{eq : sum_tm}),
but, as we will see below, it will be convenient to have two $\epsilon$ parameters.
In any case, we can rewrite (\ref{eq : sum_tm}) as:
\begin{equation} \label{eq : sum_tm1}
P(\omega_T^{0}= + ) \leq \sum \limits_{t_m = 0}^T S_{\gamma_t} (t_m, \epsilon, \epsilon)
 \end{equation}

Inserting the bound \eqref{eq : kappa} on $\kappa$ in the last line of \eqref{eq : def_S} gives (with $L = k( n_{t_m-1}-e_{t_m-1})$ and summing over $j=L-n_{t_m}= r_{t_m-1}$)

\begin{align} \label{eq : calc_S}
C^L \sum \limits_{n_{t_m} = 1}^{L}
		 \frac{
		      \kappa_{n_{t_m}}^{n_{t_m-1}-e_{t_m-1}}
		    \epsilon_0^{n_{t_m}}}
		    {\gamma_{t_{m-1}}^{r_{t_m-1}}}
	  \leq &
     C  ^ {L }\epsilon_0 \sum \limits_{j = 0}^{L-1}
 		\binom{L-1}{j}
       \frac{L !}{(L-j)!}
	    \frac{\epsilon_0^{L-j-1}}{(\gamma_{t_m-1})^{j}}
	     \nonumber \\
  \leq & C\epsilon_0 \left(C \epsilon_0 + \frac{L C}{\gamma_{t_m-1}}   \right)^{L-1} 
  \nonumber \\
  \leq & \left(C \epsilon_0 + \frac{L C}{\gamma_{t_m-1}}   \right)^L 
 \nonumber \\
  \leq & \left(C \epsilon_0 + \frac{k^{t_m} C}{\gamma_{t_m-1}}   \right)^L
.
\end{align}
To obtain the second line, we used $\frac{L!}{(L-j)!} \leq L^j$ to express the sum as a binomial.
The last line is obtained by bounding $L$ by $k^{t_m}$, 
since the number of sites at a given time is always smaller then $k$ times the number of sites at the previous time, so that we have:
$L = k( n_{t_m-1}-e_{t_m-1}) \leq k n_{t_m-1}\leq k^2 n_{t_m-2} \cdots \leq k^{t_m}$.
We obtain:
\begin{align*}
 S_{\gamma_t} (t_m,  \epsilon, \epsilon_0)
    \leq &
 \sum \limits_{\substack{ (n_t)_{t \leq t_m - 1}  \\ (e_t)_{t \leq t_m - 2}}} 
    \prod \limits_{t = 0}^{t_m - 2} 
		\frac{
		\binom{n_t}{e_t}
		\kappa_{n_{t+1}}^{n_t - e_t} 
		\epsilon^{e_t} }
		 {\gamma_t^{r_t}}  \cdot \\
  & \quad \quad
	   \cdot \sum \limits_{e_{t_m-1} = 0}^{n_{t_m-1}}
	    \binom{n_{t_m-1}}{e_{t_m-1}}
	    \epsilon^{e_{t_m-1}}
	    \left(C \epsilon_0 + \frac{k^{t_m} C}{\gamma_{t_m-1}}   \right)^{k( n_{t_m-1}-e_{t_m-1})}
\\
    = &     
	S_{\gamma_t}\left(t_m-1, \epsilon, \epsilon_1 \right),
\end{align*}
where we use
\begin{equation} 
\sum \limits_{e_{t_m-1} = 0}^{n_{t_m-1}}
	    \binom{n_{t_m-1}}{e_{t_m-1}}
	    \epsilon^{e_{t_m-1}}
	    \left(C \epsilon_0 + \frac{k^{t_m} C}{\gamma_{t_m-1}}   \right)^{k( n_{t_m-1}-e_{t_m-1})} = \epsilon_1^{n_{t_m-1}}
 \nonumber
 \end{equation} 
with
\[
 \epsilon_1 = h_{t_m} (\epsilon_0) = \left(C \epsilon_0 + \frac{k^{t_m} C}{\gamma_{t_m-1}}   \right)^k + \epsilon.
\]
So, we bound  the sum $S_{\gamma_t} (t_m,  \epsilon, \epsilon_0)$ by a similar sum, $S_{\gamma_t}\left(t_m-1, \epsilon, \epsilon_1 \right)$, but with a ``renormalized" $\epsilon$ for the last time in the sum.
We would now like to iterate this transformation $t_m$ times. 
This is not straightforward, since time  enters also in the parameters of $h_t$.

\begin{figure}
\begin{center}

\begin{tikzpicture}[scale = 0.9]
       \begin{axis}[width = 7cm, height = 7cm, ymax = 1, ymin = 0, xmax = 1, xmin = 0,
   	   ytick = { 0.05 },
 	   yticklabels = { $h_t(0)$},
	    xtick = {0,0.25,1},
	    xticklabels = {$0$, $\tilde{x}_t$, $1$}
	    ]
      \newcommand \eps {0.05}
      \newcommand \Aa {0.0}
      \newcommand \C {2}
	  \addplot[ultra thick, light-gray][domain = 0:1]{x};
          \addplot[][domain = 0:1]{ \eps+ \C*(x+ \Aa)^2};
            \addplot[dashed][domain = 0.1:0.4]{x-0.075};
  \end{axis}
\end{tikzpicture}
\end{center}
\caption{If condition \eqref{eq : cond_iteration} is satisfied, 
$h_t$ has two fixed points and $\epsilon$ stays small through the iterations provided its initial value is small enough.}
\label{fig : iteration}

\end{figure}
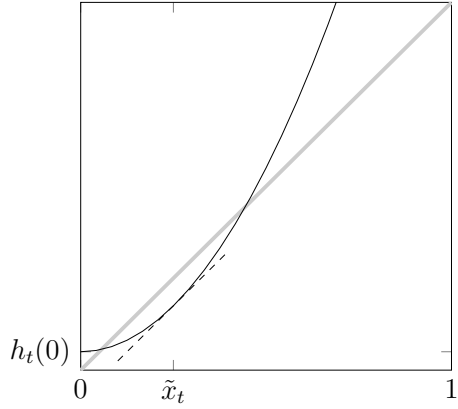

If, for some $t$, $h_t$ has the shape sketched in figure \ref{fig : iteration}, 
we have that $x \leq \tilde x_t $ implies $h_t (x) \leq h_t (\tilde x_t) \leq \tilde x_t$, 
where $\tilde x_t$ is the point such that $h'_t (\tilde x_t) = 1$.
One checks that  $h_t (\tilde x_t) \leq \tilde x_t$ when:

\begin{equation*} 
\epsilon \leq \frac{1}{C}\left( \frac{1}{C k}\right)^ \frac{1}{k-1} -\left( \frac{1}{C k}\right)^ \frac{k}{k-1} 
	      - \left(\frac{k^{t}}{\gamma_{t-1}} \right).
\end{equation*}
To have the above inequality at all times, $\gamma_t$ should grow exponentially with $t$.
The simplest possibility is to take $\gamma_t = gk^t$ and choose $g$ and $\epsilon$ such that

\begin{equation}
 \label{eq : cond_iteration}
\frac{k}{g} \leq \frac{1}{C}\left( \frac{1}{C k}\right)^ \frac{1}{k-1} -\left( \frac{1}{C k}\right)^ \frac{k}{k-1}
\quad \text{and} \quad
\epsilon \leq \frac{1}{C}\left( \frac{1}{C k}\right)^ \frac{1}{k-1} -\left( \frac{1}{C k}\right)^ \frac{k}{k-1} 
	      - \frac{k}{g}.
\end{equation}

In this case all the iterates of $\epsilon$ stay smaller than 
$\tilde{x}  \equiv
\frac{1}{C}\left( \frac{1}{C k}\right)^ \frac{1}{k-1}- \frac{k}{g}$, since, with our choice of  $\gamma_t $,  $\tilde{x}_t$ does not depend on $t$.

We now have the bound
\begin{align*} \label{eq : maj_S}
 S_{\gamma_t} (t_m, \epsilon, \epsilon) &\leq S_{\gamma_t} (0,\epsilon ,\epsilon_{t_m}) = \epsilon_{t_m} 
  \leq \tilde{x},
\end{align*}
since the only graph with $t_m = 0$ consists of an error site only.

However, this bound is  not sufficient to control the sum over $t_m$ in  \eqref{eq : sum_tm1}.

\subsection{Summing over $t_m$}
In order to obtain a useful bound to control the sum  in  \eqref{eq : sum_tm1}, we define $\epsilon = \delta \epsilon'$ and $\gamma_t = g {p}^{t+1} k^t \equiv p^{t+1} \gamma'_t $.
$\epsilon'$ and $\gamma_t'$ will replace $\epsilon$ and $\gamma_t$ in the previous calculations 
and will be assumed to satisfy \eqref{eq : cond_iteration}, 
while the additional factors $\delta < 1$ and $p > 1$ will make the last sum converge. We will also assume that $\delta$ is small enough so that $p \delta^{k-1} \leq 1 $.

First, observe that if a graph corresponding to a term in the sum (\ref{eq : sum_tm}) contains a tree of height $h$ as a subgraph, then that tree contains at least $h (k-1) +1$ error sites:
indeed, for $h=1$, we have just $k$ lines going  from one site at time $T-t $ to an error site at time $T-t-1$. Next, observe that we can replace  a tree of height $h$ by one of height $h-1$
by replacing, at least once (but possibly more times), $k$ error sites  at time $T-t-1$ by one  error site at time $T-t$, so that the tree of height $h$ has at least $k-1$ error sites more than the tree of height $h-1$. This proves  inductively our claim.

Next, we see that each graph  contains a  tree of some height $h$ between $0$ and $t_m$, where $h$ is the largest
  $h$ such that $r_{t_m}= \hdots = r_{t_m-h} = 0$ ($r_{t_m} = 0$ by definition of $t_m$). Indeed, since $r_t= k (n_{t}- e_{t}) - n_{t+1}$, $r_t=0$ means that all the $k (n_{t}- e_{t})$ lines going from time $T-t$ to time $T-t-1$ go to different sites,
  so that the part of the graph below time $T-t_m + h$ is  made of disjoint trees, at least one of which is of height $h$.
 If $h = 0$, it means $r_{t_m -1} \neq 0$ and, with our redefinition of   $\epsilon$  and $\gamma_t$ , we get an additional factor of at least $\frac{\delta}{p^{(t_m -1) +1}}=\frac{\delta}{p^{t_m }}$, since there is at least one error at $t_m$, and one 
 extra factor of $p^{-(t+1)}$ coming from one of the factors $\gamma_t^{-1}$ in 
(\ref{eq : sum_tm}), for $t=t_m-1$, since  $ r_{t_m-1} \geq 1$.
If $h \neq 0$, 
we similarly get an additional factor smaller than $ \frac{\delta^{h(k-1) +1}}{ p^{t_m -h } }$: the factor $\delta^{h(k-1) +1}$ comes from the error sites and $p^{-(t_m -h)}$ from one of the factors $\gamma_t^{-1}$ in 
(\ref{eq : sum_tm}), for $t=t_m-h-1$, since  $ r_{t_m-h-1} \geq 1$.
Since we have chosen $p \delta^{k-1} \leq 1 $, we have 
 $ \frac{\delta^{h(k-1) +1}}{ p^{t_m -h } }\leq  \delta  p^{-t_m}  $.

This proves:
\begin{align*}
S_{\gamma_t} (t_m, \epsilon, \epsilon)
	\leq    S_{\gamma'_t}(t_m, \epsilon',\epsilon')    \delta  p^{-t_m},  
\end{align*}
and therefore:

\begin{align*}
P(\omega_T^{0}= + ) &
             \leq \sum \limits_{t_m = 0}^T S_{\gamma'_t} (t_m, \epsilon', \epsilon')  \delta  p^{-t_m}\\
      & <  \sum \limits_{t_m = 0}^T \tilde{x}
		    \delta  p^{-t_m}						\\
      &< \delta \frac{p}{p-1} \tilde{x}.
\end{align*}

This bound can be made arbitrarily small as $\delta$ decreases, 
so this proves that there is a phase transition in the intermediate model.

\section*{Acknowledgments} This work is partially funded  by the Belgian Interuniversity Attraction Pole, P7/18.

\end{document}